\documentclass[letter]{jfm}

\usepackage{graphics}
\usepackage{epsfig}

\ifCUPmtlplainloaded \else
  \checkfont{eurm10}
  \iffontfound
    \IfFileExists{upmath.sty}
      {\typeout{^^JFound AMS Euler Roman fonts on the system,
                   using the 'upmath' package.^^J}
       \usepackage{upmath}}
      {\typeout{^^JFound AMS Euler Roman fonts on the system, but you
                   dont seem to have the}
       \typeout{'upmath' package installed. JFM.cls can take advantage
                 of these fonts,^^Jif you use 'upmath' package.^^J}
       
      }
  \else
    
  \fi
\fi

\ifCUPmtlplainloaded \else
  \checkfont{msam10}
  \iffontfound
    \IfFileExists{amssymb.sty}
      {\typeout{^^JFound AMS Symbol fonts on the system, using the
                'amssymb' package.^^J}
       \usepackage{amssymb}
       \let\le=\leqslant  \let\leq=\leqslant
         
      }{}
  \fi
\fi

\ifCUPmtlplainloaded \else
  \IfFileExists{amsbsy.sty}
    {\typeout{^^JFound the 'amsbsy' package on the system, using it.^^J}
     \usepackage{amsbsy}}
    {}
\fi

\newsavebox{\astrutbox}
\sbox{\astrutbox}{\rule[-5pt]{0pt}{20pt}}

\title[Chaotic Dripping Faucet]{Hydro-dynamical models \\
for the chaotic dripping faucet}

\author[P. Coullet, L. Mahadevan and C.S. Riera]
{P.\ns C\ls O\ls U\ls L\ls L\ls E\ls T$^1$
,\ns
L.\ns M\ls A\ls H\ls A\ls D\ls E\ls V\ls A\ls N$^{2,3}$\break
\and C.\ns S.\ns R\ls I\ls E\ls R\ls A$^{2,3}$}

\affiliation{$^1$INLN, 1361 route des Lucioles, 06560 Valbonne, FRANCE\\[\affilskip]
$^2$DAMTP-CMS, Wilberforce Road, Cambridge CB3 0WA, United Kingdom\\[\affilskip]
$^3$Present address, DEAS, Harvard University, 29 Oxford Street, Cambridge MA 02138 USA}

\pubyear{1996}
\volume{538}
\pagerange{119--126}
\date{Feb 2004 and in revised form Aug 2004}
\begin{document}

\maketitle

\begin{abstract}

We give a  hydrodynamical explanation for the chaotic behaviour of a dripping faucet using the results  of the stability analysis of a static pendant drop and a proper orthogonal decomposition (POD) of the complete dynamics. We find that  the only relevant modes are the two classical normal forms associated with a Saddle-Node-Andronov bifurcation and a Shilnikov homoclinic bifurcation. This allows us to construct a hierarchy of reduced order models including maps and ordinary differential equations which are able to qualitatively explain prior experiments and numerical simulations of the governing partial differential equations and provide an explanation for the complexity in dripping. We also provide a new mechanical analogue for the dripping faucet and a simple rationale for the transition from dripping to jetting modes in the flow from a faucet.

\end{abstract}

\section{Introduction}

Almost since the beginning of the modern revolution in nonlinear dynamics and chaos, the dripping faucet has served as a paradigm of chaotic dynamics (\cite{Shaw}). To explain the transition to chaos, various ad-hoc mechanical models  based on variable-mass and spring systems have been used to derive return maps and Poincar\' e sections for the time between droplet emissions. Nearly twenty years on, this empirical approach still continues (see (\cite{kiyono})  for a recent example along with a review of earlier work), while the connection to hydrodynamics remains tenuous. In contrast, over the last decade, the hydrodynamical approach has been used with great success in studying the pinch-off of a single drop using a combination of theory, numerical simulation and experiment (for a review see (\cite{eggers})).   Thus it is natural to bring this understanding of the hydrodynamics  of drop emission to bear upon the complex dynamics of dripping and understand its origins. Two recent attempts in this direction are  \cite{fuchikami} and \cite{Basaran}  who simulate the chaotic dynamics of dripping using simplified lubrication-like approximations of the Navier-Stokes equations.  However, the mechanistic origins of the complexity in dripping still remain elusive.

Our goal in this paper is not as much to match the images of chaotic dripping seen in experiments or numerical simulations as to see how  we might qualitatively explain what seems to be a complex dynamical process as simply as possible.  With this in mind, we derive a series of models culminating in a mechanical analogue, all of which capture the salient features of periodic and chaotic dripping and jetting, thus bringing a focus on the physical mechanisms responsible while providing a guide to further work on related problems.  Some preliminary results of this nature were first announced in a conference proceedings (\cite{PTP}) a few years ago. In \S 2, we revisit the equations of equilibrium for a static pendant drop. In \S 3, we describe a Lagrangian model based on lubrication theory for the evolution of the drop and use it to carry out a linearized analysis of the static drop shapes as a function of their volume. In addition, we also present solutions of the complete dynamical evolution associated with dripping for a range of flow rates and evaluate the bifurcation diagram that shows the transition to chaos.  In \S 4, we analyze these chaotic solutions using proper orthogonal decomposition and a time-delay method, both of  which suggest that a low dimensional model should suffice to explain the complexity of dripping.  In particular, we show that the dynamics can be recast in terms of a hierarchy of low-dimensional models, i.e. coupled ordinary differential equations and maps, in \S5  which are qualitatively consistent with prior experiments and  yield a simple physically consistent picture of the complex dynamics in terms of well known concepts in dynamical systems. Guided by these models, we propose a new mechanical analogue of the dripping faucet in \S6. We conclude the paper with a brief discussion in \S7 of how our models can rationalize the  transition from dripping to jetting.

\section{Equations of equilibrium}

We start with  the case where the flow rate is very small, so that a drop remains attached to the faucet until its volume exceeds a threshold $V_c$. For a faucet of radius $R$ that is sufficiently small, drops with a volume $V<V_c$ are stable and axisymmetric (\cite{padday}) which case we will restrict ourselves to.  The shape of such a pendant drop is determined by minimising the sum of its gravitational and surface energy subject to the constraint of constant volume. Alternatively, we can write the dimensionless equation for the balance of forces normal to the interface as 

\begin{equation}
\displaystyle\frac{d\theta}{ds}  -  \displaystyle\frac{\cos\theta}{r}= -z,
\label{forcebalance}
\end{equation}

where the drop interface $(r(s),z(s))$ is determined by the kinematic conditions 

\begin{equation}
\left \{
\begin{array}{rcl}
\displaystyle\frac{dz}{ds}&=&-\cos\theta\nonumber \\ 
\displaystyle\frac{dr}{ds}&=&\sin\theta\label{kinematic}
\end{array}
\right.
\end{equation} 

\begin{figure}
\begin{center}
\includegraphics[width=10cm]{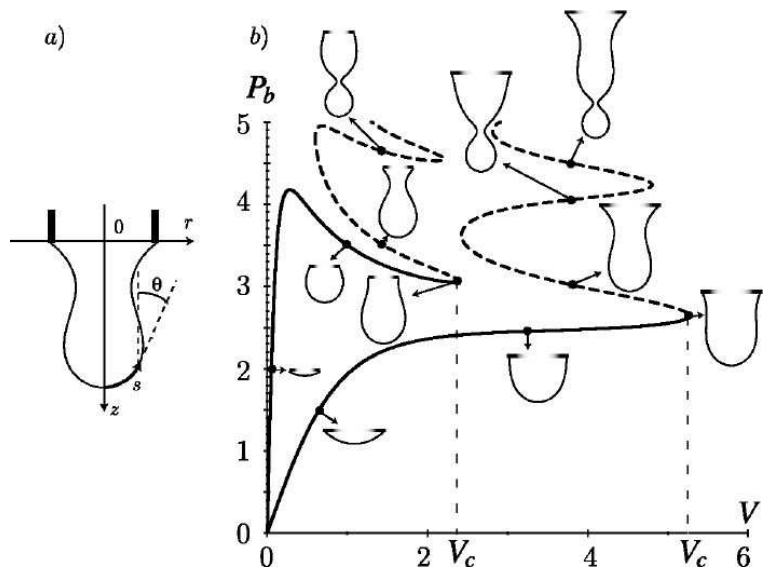}
\caption{a) Schematic of the static drop. b) The dimensionless hydrostatic pressure at the bottom of the drop $P_{b}$ as  a function of the dimensionless volume $V$ of the drop for different values of the dimensionless faucet radius $R=0.5$ (left curve) and $R=1.0$ (right curve). Solutions of (\ref{forcebalance}) and (\ref{kinematic}) yield the $P-V$ curves shown, with the the solid line corresponding to stable static drops and the dashed line corresponding to unstable stationary drops, }
\label{volvspb}
\end{center}
\end{figure}

Here the variables $r$, $s$, $\theta$ and $z$ are defined in Fig. (\ref{volvspb}a). The  characteristic scales are :  $l_0=\sqrt{\Gamma/g\rho}$ for  length,  $m_0=\rho l_0^3$ for mass,  $t_0=(\Gamma/\rho g^3)^{1/4}$ for  time, where $\Gamma$, $\rho$, $g$ are the surface tension, density of the fluid and gravity respectively. For water at $20 ^0C$, $l_0=0.27cm, m_0=0.020g, t_0 =0.017 s$. The boundary conditions at the bottom of the drop are $r(0) = 0$, $\theta(0) = \pi/2$ and $z(0) = P_b / \rho g$,  where $P_b$ is the unknown hydrostatic pressure at this position and serves as a control parameter that describes the family of stationary drops. Choosing a value for $P_b$, we integrate  (\ref{forcebalance},\ref{kinematic})  as an initial value problem until $r=R$ ($R=cst $) and find the corresponding drop volume $V=\int \pi r^2 dz$.  For a 
given $P_{b}$ when $R<1$ the condition $r=R$ may be satisfied for up to three different drop volumes and lengths (\cite{riera}). These  pendant drops are shown in Fig. (\ref{volvspb}) for different faucet radii $R$. However, the only stable stationary drop corresponds to the branch starting at the origin and ending at the first turning point (\cite{padday}) which denotes the critical drop volumes $V_{c}$ for which the weight of the drop is just balanced by the capillary forces. The corresponding shapes have a waist where the dynamic instability associatedÊwith eventual pinch off first starts.  

\section{Equations of motion} 

 To understand the mechanism of this linear instability dynamically, we consider the hydrodynamical equations linearized about a stationary solution. Instead of using the complete Navier-Stokes equations for Eulerian one-dimensional lubrication theories (\cite{dupont}), we simplify the analysis by using a new lubrication model embodied in a Lagrangian approach for the fluid (\cite{fuchikami}). The inherent assumptions in this are the following: (a) the drop remains axisymmetric during its motion, (b) the radial component of the fluid velocity is negligible compared to the axial component which depends only on $z$, (c) there is no overturning of the interface $r(z)$. These assumptions are asymptotically valid for slender drops of large viscosity, but recent simulations of the resulting low-order equations (\cite{fuchikami}) have shown good agreements with experiments even for fairly squat drops of low viscosity. The above assumptions lead to the conclusion that there is no exchange of fluid between neighboring horizontal slices of the drop, so that the volume of a slice is constant during the motion and can be treated as a Lagrangian variable. This yields a formulation  that is essentially equivalent to earlier Eulerian description. Explicitly, the volume between the planes $z_b(t)$ and $z_a(t)$ is
 
\begin{equation}
\xi(z_a,z_b,,t)=\int_{z_a(t)}^{z_b(t)}\pi r(\zeta,t)^2d\zeta
\end{equation}

where $r$ is the radius of the drop. In terms of the Lagrangian variable $\xi(z_a,z_b,t)$, we can write the kinetic energy $E_{kin}$, potential energy $U_g$  and surface tension energy $U_\Gamma$ of the system as 

\begin{eqnarray}
\left \{
\begin{array}{lcl}
E_{kin} &=& \frac{\rho}{2}\int_0^{\xi_0(t)}\left(\frac{\partial z(\xi,t)}{\partial t}\right)^2d\xi \\
U_g &=& -\rho g\int_0^{\xi_0(t)}z(\xi,t)d\xi \\
U_\Gamma &=&\Gamma\int_0^{\xi_0(t)}\sqrt{4\pi z'+\frac{(z'')^2}{(z')^4}}d\xi
\end{array}
\right.
\end{eqnarray}

Here $\xi_0(t)$ is the total volume of the drop at  time $t$, and a prime corresponds to a partial derivative with respect to the Lagrangian variable $\xi$. Then we can write the Lagrangian of the system as 

\begin{equation}
{\mathcal L} = E_{kin} - U_g - U_\Gamma
\end{equation}

The effect of viscosity is then expressed using the Rayleigh dissipation function

\begin{equation}
\dot E_{kin} = -3\eta\int_0^{\xi_0(t)}\left(\frac{v'(\xi,t)}{z'(\xi,t)}\right)^2d\xi
\end{equation}

Then, Lagrange's equation for the system is

\begin{equation}
\frac{d}{dt}\frac{\partial\mathcal{L}}{\partial v} = \frac{\partial\mathcal{L}}{\partial z} +
 \frac{1}{2}\frac{\partial\dot E_{kin}}{\partial v}
\end{equation}

\begin{figure}
\includegraphics[width=12cm]{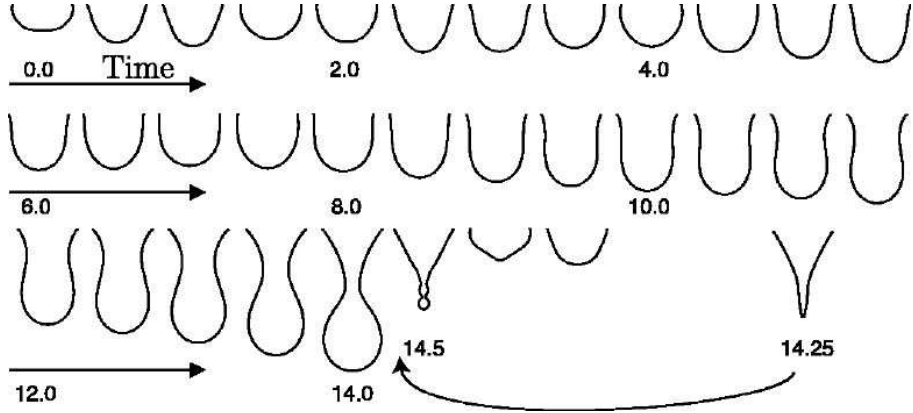}
\caption{The evolution of the drop as a function of time computed by solving (\ref{disc}).   The snapshots are separated by $0.5$ unit of normalised time, unless marked otherwise. Parameters values for the simulation are $R=1,\eta=0.02,g=1,\gamma=1,v_{0}=0.1$.}
\label{snapshot}
\end{figure}

\begin{figure}
\includegraphics[width=12cm]{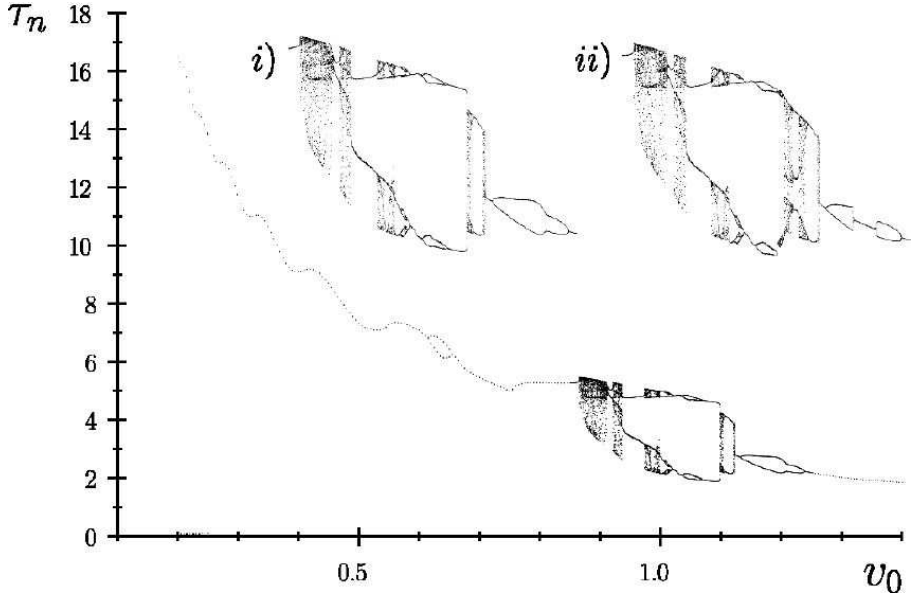}
\caption{Time interval between drops as a function of the exit velocity $v_{0}$, obtained by solving (\ref{disc}), with $R=0.5,\eta=0.02,g=1,\gamma=1$. The qualitative features that are apparent are the oscillation of the time between drops as a function of the exit velocity $v_{0}$, the single period-2 oscillations before the chaotic region and finally the transition to chaos via period doubling and via a boundary crisis. All these features are robust functions of  parameters, i.e. they persist for a range of values. In the inset (i), we show the bifurcation diagram associated with increasing flow rate and in (ii) the corresponding diagram with decreasing flow rate; we observe that there is hysteresis.}
\label{fullsimu}
\end{figure}

For the purposes of computation we follow the method of \cite{fuchikami} and discretize the Lagrangian spatially so that the drop is effectively sliced into $N$ disks; each disks is characterised by 3 variables : the position $z_i$, the velocity $v_i=\frac{\partial z_i}{\partial t}$ and the mass $m_i$. A disk is divided in two when its relative thickness (ratio of its height to its width) exceeds a threshold ($0.05$ in a typical simulation) to keep it slender. This occurs, either because of the added volume at the faucet or the change in the shape of the drop in the necking area. Two consecutive disks are combined when their relative thickness falls below a different threshold ($0.075$). The splitting and merging are done so as to conserve volume and momentum\footnote{Other quantities are not conserved as accurately, as in \cite{fuchikami}}. This auto-adaptive mesh leads to a variation of $N$ between $50$ and $4000$ during a typical simulation (see Fig. \ref{snapshot} and Fig. \ref{fullsimu}).   

The discretized Lagrangian then yields  the $N$ equations of motion 

\begin{equation}
\frac{d}{dt}\frac{\partial\mathcal{L}}{\partial v_i}=\frac{\partial\mathcal{L}}{\partial z_i}+
\frac{1}{2}\frac{\partial\dot E_{kin}}{\partial v_i}, ~~~i=1,N. 
\label{disc}
\end{equation}

These equations of motion are then integrated using a fifth-order Runge-Kutta method with an adaptive timestep that is based on the local truncation error (see Fig. \ref{snapshot}). The initial condition is either a stable stationary solution of the equation of equilibrium (\ref{forcebalance},\ref{kinematic}) or a hemisphere.

\begin{figure}
\includegraphics[width=12cm]{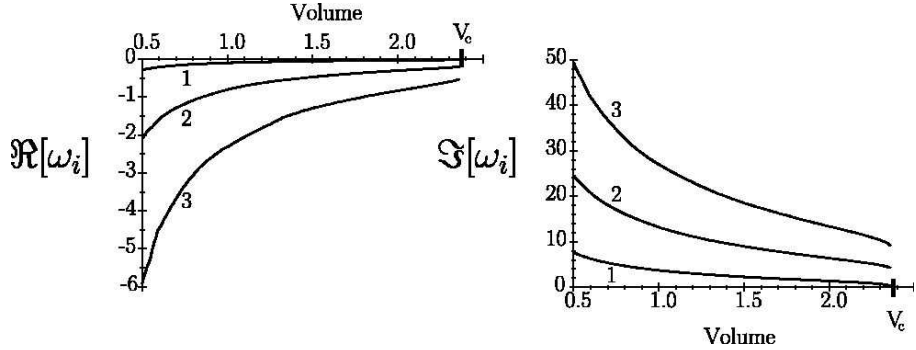}
\caption{ Evolution of the real and imaginary part of the largest three eigenvalues $(\Re [\omega_i],\Im [\omega_i])$ as a function of the scaled drop volume $V$ shows that $\Re [\omega_1] \sim 0.1\Re [\omega_2]$ ($R=0.5,\eta=0.02,g=1,\gamma=1$).}
\label{eigenvaluea}
\end{figure}

\begin{figure}
\begin{center}
\includegraphics[width=8cm]{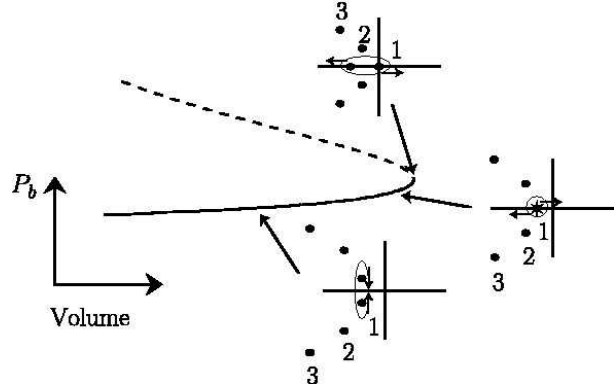}
\caption{Sketch of the location of the  first few eigenvalues when  $V \le V_{c}$. The first "nose" of the $P_b - V$ curve corresponds to the collision between the  unstable "saddle" and stable "node" solutions and leads to a saddle-node bifurcation.}
\label{eigenvalueb}
\end{center}
\end{figure}

To determine the stability of the static solution we use the solutions determined in \S2  via a shooting method  (since there are no analytic solutions for the static shape) and substitute them into the  linearized equation of motion (\ref{disc}) in the neighbourhood of the stationary solutions of (\ref{forcebalance},\ref{kinematic})  and  determine the eigenvalues $\omega_i, i=1, N$ of the resulting system. The value of the real part and imaginary part of the $3$ largest eigenvalues are plotted in Fig. (\ref{eigenvaluea}). When $V< V_c$, $\Re [\omega_i] <0$, so that these drops are stable. As $V \rightarrow V_c$ two complex  conjugate eigenvalues become real by colliding on the real axis (Fig. (\ref{eigenvalueb})) corresponding to the critically  damped oscillator. As $V$ increases further, the eigenvalues split; but remain on the real axis. One of the eigenvalues then moves away from the imaginary axis and the other  moves  towards it eventually reaching the origin when $V=V_c$. This is consistent with Fig. (\ref{eigenvalueb}) where we show the volume of the drop as a function of $P_{b}$. The collision of two branches of stationary solutions one stable and one unstable and the disappearance of the stationary solution when $V>V_c$ is a characteristic of a saddle-node bifurcation.

At low flow rates, the shape of the drop is always close to that of a stationary drop. The spectrum of the linearized stable solutions with $V \le V_{c}$, as shown in Fig. (\ref{eigenvaluea}) indicates that only two modes are dynamically relevant because the others are rapidly damped. These modes correspond to a) an oscillatory damped mode given by the largest two complex conjugate eigenpairs  and b) a Saddle-Node bifurcation (Fig. (\ref{eigenvalueb})). To understand the dynamics of drop emission and relaxation, we now turn towards the numerical solution of the equations of motion (\ref{disc}) with the eventual goal of justifying a low order approximation for this complex process.

\section{Galerkin proper orthogonal decomposition}

Proper Orthogonal Decomposition (POD) (\cite{holmes}) allows us to project a a high/infinite dimensional system onto a finite number of basis functions or modes. This may be done using the data from the direct numerical simulations, time series etc. and allows us to try and understand the qualitatively important features of a complex process.  The optimality of POD derives from the fact that a truncated POD describes typical members of the ensemble better than any other decomposition. 
We use POD to determine if a low mode approximation of the dynamics of dripping is valid. As an example we carry out a POD in a regime with period-2 dripping ($v_{0}=1.025 \ll l_{0}/\tau_{0}$, flow rate $q=\pi R^{2}v_{0}$ and  $R=0.5$) where there is  noticeable difference in the time required for the formation of large and small drops, allowing us to clearly distinguish the two periods. 

The shape of the drop $r(z,t)$ is recorded periodically and a regular spaced mesh based on the maximum drop length is used to interpolate the typically irregular mesh. If a point in the interpolated regular mesh is outside the drop we set the radius $r=0$. This mesh is used to compute both the average shape of the drop and the variation from this shape. This procedure allows us to follow the variance of the shape and leads to a rectangular matrix $M$ of size $m\times n$ corresponding to $m$ snapshots of the variation of the drop shape on a regular spaced mesh containing $n$ points. The POD of $M$ is a decomposition of the form $M=UDV^{T}$, where $V$ is an $n\times n$ matrix and $U$ an $m\times n$ matrix. The columns of the matrix $V$ form an orthonormal basis corresponding to the spatial modes of the drop. The columns of the matrix $U$ also from an orthonormal basis and correspond to the normalised projection of each snapshot onto the spatial modes of the POD, i.e. $MV=UD$ where the diagonal matrix $D$ corresponds to the singular values. 

\begin{figure}
\includegraphics[width=13cm]{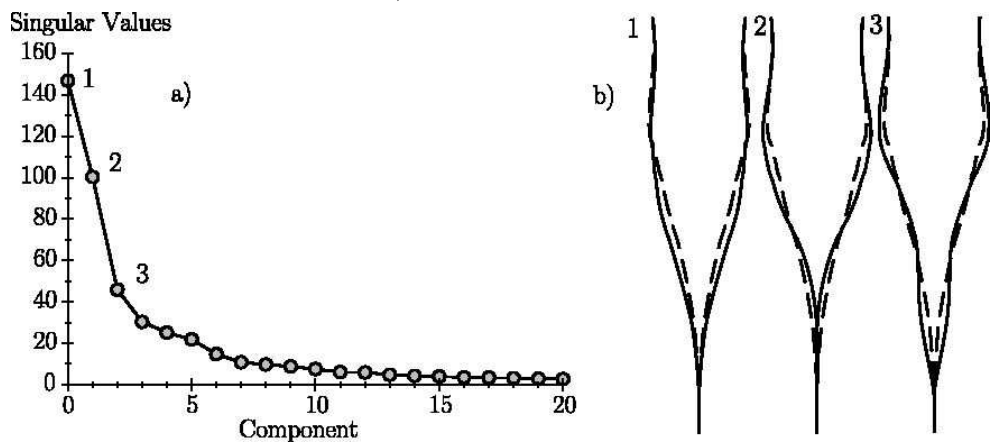}
\caption{A proper-orthogonal decomposition (POD) of the numerical  simulation of the hydrodynamical equations (\ref{disc}) for the dripping faucet ($R=0.5$, $\eta = 0.02$ and $v_{0} = 1.025$). a) Singular values  are plotted for each "mode". b) The first three modes from the POD  superposed on the average drop shape: 1 corresponds to $V_{1}+r_{a}$, 2 corresponds to $V_{2}+r_{av}$ and 3 corresponds to $V_{3}+r_{a}$(the dashed line corresponds to the interpolated average drop shape $r_{a}$) computed as described in the text. Here $V_{i}$ corresponds to the $ith$ column of the matrix POD as indicated in the text. We see that just the first three modes capture much of the dynamical behavior of the faucet.}
\label{pod}
\end{figure}

\begin{figure}
\includegraphics[width=10cm]{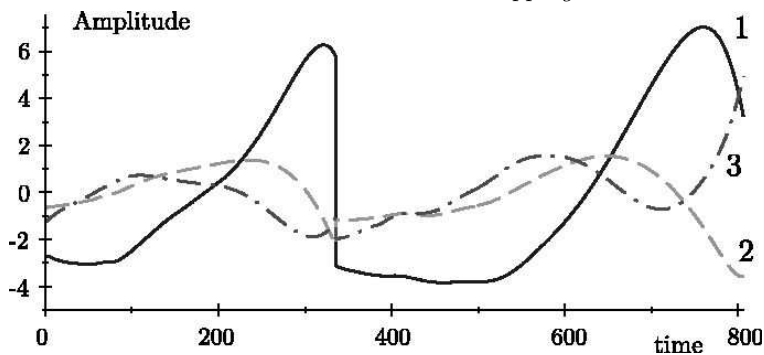}
\caption{ Projection of the dynamics embodied in the solution to (\ref{disc}), onto the first three POD modes. The solid line corresponds to the first mode $U_{1}D_{1}$, short dashed line the second $U_{2}D_{2}$ and long dashed and dotted line the third $U_{3}D_{3}$, with $U_{j}$ being the $jth$ column and $D_{j}$ being the diagonal entry in the matrix POD as indicated in the text.}
\label{podtemp}
\end{figure}

In Fig. \ref{pod} we show that the singular values of the first two modes are distinct from the other  clustered modes for the particular choice of parameters indicated earlier. This separation of the singular values is strongly dependent on the scaled viscosity $\eta$; as $\eta$ decreases all the singular values cluster together. The first mode ( 1 in Fig. \ref{pod})  is associated with the process of drop growth and emission  (Fig. \ref{podtemp} solid line). The second mode  (2 in Fig. \ref{pod}) corresponds to the largest damped oscillatory mode (Fig. \ref{podtemp} short dashed line). Since the higher modes are rapidly damped, we see that the dynamics of the dripping faucet can be well approximated using only the first two modes of the basis obtained using POD in this period-2 regime. Similar results are obtained in different regimes.

We note that {\em a priori}  there is no direct connection between the linearized modes and the POD modes. Here we distinguish the two uses of POD. Traditionally, it is used a method of simplifying the complex dynamics of high dimensional system. Here our goal is slightly different; what we want to show is that a low dimensional description is possible, and in following we use the linearly stable modes to construct a hierarchy of low dimensional models.

\begin{figure}
\includegraphics[width=10cm]{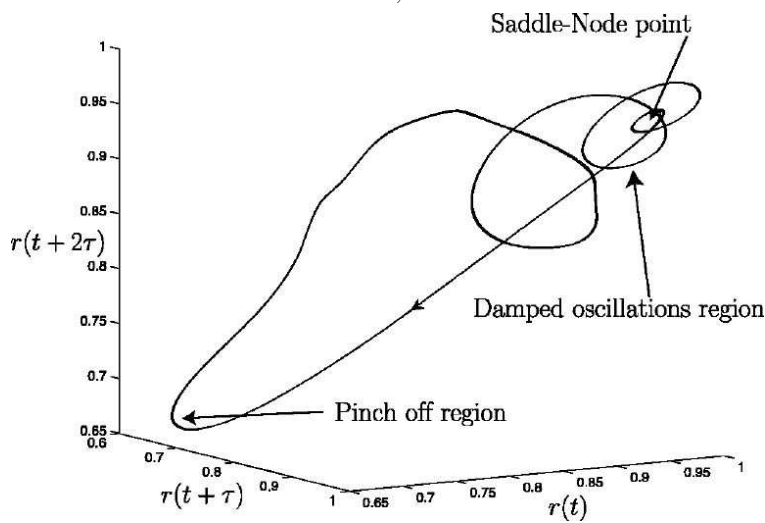}
\caption{ Reconstruction of the flow by the time delay method obtained by solving ({\ref{disc}}) numerically with parameter values corresponding to those for Fig \ref{snapshot}. The radius of the drop $r(t+i\tau)=r(0.5,t+i \tau), i=1,2,3$   so that it remains continuous during the process, and the time delay $\tau = 1.0$. We observe a long excursion followed by a damped oscillations before the orbit returns to the neighbourhood of the saddle-node point. The arrow indicates the direction of the flow. }
\label{phasespace}
\end{figure}

We can also reconstructed the phase space of the dynamics by using the time delay method (\cite{ruelle}). We use the radius of the drop at a given position above the usual location of the pinch-off so that the variable remains continuous during the process. We choose different delays $T_{0}=0,T_{1},...,T_{N-1}$ to define our new variables $r_{k}=r(t+T_{k})$ which gives us a $N$-dimensional system. The minimum dimension $N$ of the system is chosen so that the flow doesn't cross itself in the phase-space and the delays are chosen somewhat arbitrarily in order to obtain a reasonable projection. We find that $N=3$ is sufficient to capture the dynamics suggested by both the linear analysis and the POD. In the phase-space generated by the time-delay method shown in (Fig. (\ref{phasespace}) we see two qualitatively different regions: a large excursion corresponding to the dynamics that lead to  drop  pinch-off, and a much more compact region corresponding to the damped oscillations following the pinch-off event, that eventually leads the orbit to the neighbourhood of the saddle-node area whence it escapes again. 

\section{Low dimensional models}

We now consider the dynamics of dripping using simplified models. The time scale for the formation of a pendant drop is $\tau_f\sim\frac{R}{v_0}$. Once the volume of the pendant drop is $V_c$, it becomes unstable and  pinches off in finite time. This process occurs in a time $\tau_n\sim R\sqrt{\rho R/\Gamma}$   (\cite{clanet}) which is much shorter than the time for a new drop to form, and independent of the flow rate. Therefore the dynamics of dripping will not be affected by the details of drop pinch-off.

After pinch-off, the remaining liquid recoils due to capillary forces, oscillating with a characteristic frequency $f \sim \sqrt{\Gamma/(\rho\mathcal{V}})$ which varies with the volume $\mathcal{V}$ and shape of the residual drop. For a given flow rate the volume of the pendant drop grows steadily,  and this frequency gradually decreases even as the oscillations are damped out by viscous fluid motions at a rate $1/\tau_d\sim\sqrt{f\eta/\mathcal{V}^{1/3}}$ (Fig. \ref{eigenvaluea}). For very small flow  rates, these oscillations are completely damped out by the time the pendant drop attains the critical volume $V_c$, so that in this case, droplets are emitted with a constant periodicity. As the flow rate is increased, these partially damped oscillations modify the onset of the instability via a saddle-node bifurcation. Equivalently, the dimensionless ratio of the filling time to the damping time $\tau_f/\tau_d$ advances or delays the onset of necking and is responsible for the variation of the periodicity (or lack thereof) of drop emission. For example, as the flow rate is gradually increased, the constant periodicity ``drop-drop'' gives way to a ``drop-drip'' scenario via a period-doubling bifurcation as follows. Once the pendant drop reaches the critical volume $V_c$, a large droplet ``drops'' leading to a highly elongated residual filament. If the flow rate is large enough so that the oscillations are not completely damped out, the next droplet will start the necking process when $V<V_c$ or $V>V_c$ depending of the direction of the oscillation close to the critical volume, so that a smaller or larger droplet ``drips''. This leads to a smaller or larger residual drop whose oscillations will be damped out much sooner or later, thereby (possibly) allowing the pendant drop reach its maximum size $V_c$ before or after it ``drops'', and so on. The temporal spacing between two drops arises from the damped oscillations close to the critical volume which enables the drop to become smaller or larger before the pinching off regime.  

\begin{figure}
\begin{center}
\includegraphics[width=8cm]{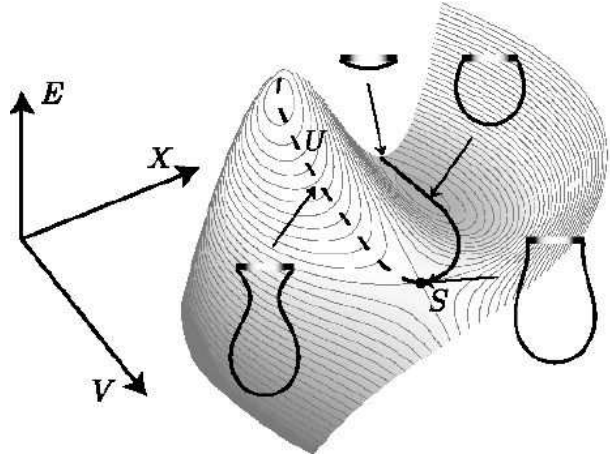}
\caption{Potential corresponding to the capillary-gravitational energy $E=a(V)X + b(V) X^3/3$ of the stable stationary pendant drop as a  function of the drop volume $V$ and the state of the drop  $X$. The stable position (solid line) for the pendant drop corresponds to the local minimum of the potential, while the dashed line denotes the unstable solution.}
\label{pot}
\end{center}
\end{figure}

Based on these observations, and the linear stability analysis consistent with the POD we build a simple low dimensional model. Let $X$ be a state variable characterising the drop  in a potential determined by the gravitational and surface energies as shown in Fig. \ref{pot}. Giving a physical interpretation of $X$ is tricky; it can be related to the center of gravity of the drop, as used for example by \cite{kiyono2}, but only indirectly because different shapes can have the same center of gravity. A better interpretation of $X$  relates it to the projection of the complete shape on a spatial mode of the drop and indicates the location of the drop relative to its stationary solutions as shown in Fig. \ref{pot} in the spirit of qualitative behaviour of  dynamical systems. In the vicinity of the minimum of the potential energy the drop oscillates stably with a frequency determined by the curvature of the potential and damping rate determined by the spectrum of the linearized equation of motion. Close to the saddle point $S$ the drop can oscillate stably or start necking, depending on the size of the perturbation. Indeed the height of the barrier represented by the unstable branch $U$ is the energy for  drop nucleation. A simple model potential characterising this is $E=a(V)X + \frac{b(V)}{3}X^{3}$, where $V$ is the drop volume. This potential evolves with the drop volume so that the stable solution disappears via a Saddle-Node bifurcation when $a(V_{c})=0$. Using these facts, we may then describe the dynamical evolution of the drop via the equations 

\begin{equation}
\left\{
\begin{array}{lcl}
\partial_{tt}X+\eta(V)\partial_{t}X & = &a(V) + b(V)X^{2} \\
\partial_{t} V & = & q 
\end{array}
\right.
\label{model}
\end{equation}

\begin{figure}
\includegraphics[width=13cm]{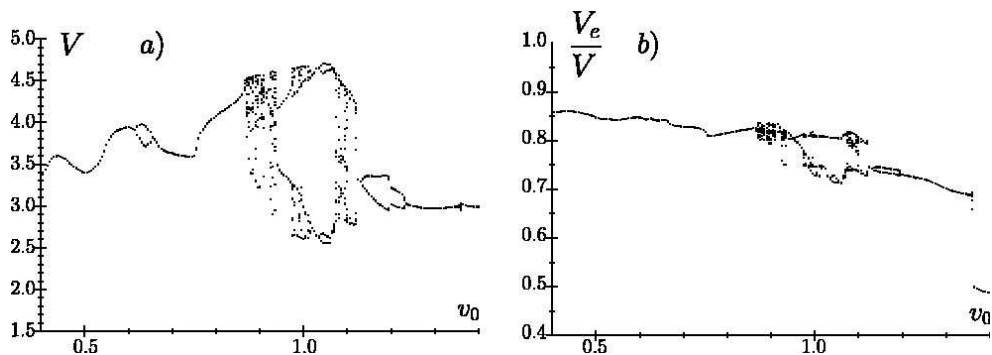}
\caption{(a) Total volume at the break up time as a function of the exit velocity $v_0$ and (b) the ratio of the  volume of the emitted drop to the total volume of the drop. To a first approximation, this ratio varies linearly with the exit velocity $v_{0}$.} \label{volvsvel}
\end{figure}

\begin{figure}
\begin{center}
\includegraphics[width=13cm]{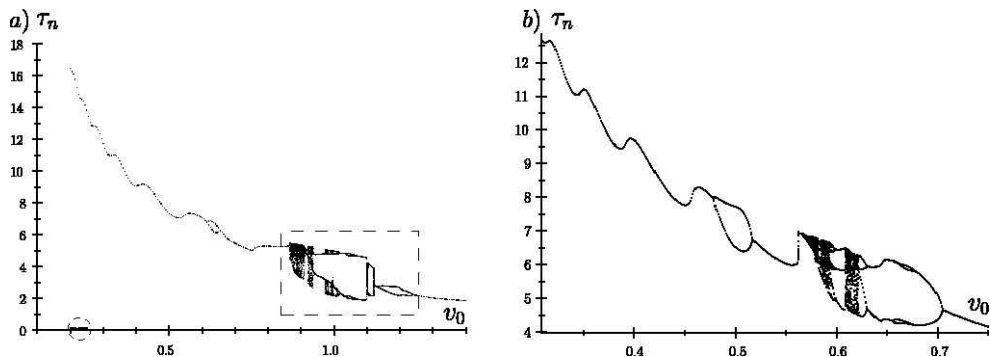}
\caption{Bifurcation diagram derived from  (a) the full hydrodynamic model (\ref{disc}) ($R=0.5,\eta=0.02,g=1,\gamma=1$).  (b)  the simple low-dimensional model (\ref{model},\ref{coef}) ($\alpha=10, \beta=0.028, \gamma=-0.106,\delta=-0.154$). $\tau_{n}$ corresponds to the time between two  drops and $v_0$ is the exit velocity. The two bifurcation diagrams are qualitatively very similar.}
\label{simplemodel}
\end{center}
\end{figure}

The first equation describes the  dynamics of the state variable $X(t)$ with a scaled damping $\eta(V)$ in the potential $E$, while the second equation quantifies the increase in the drop volume due to a flow rate $q$. The first equation of (\ref{model}) charaterizes a simple damped oscillator; the nonlinearity is the normal form for the saddle-node bifurcation while the damped oscillations arise when we account for the stable oscillations of the recoiling drop. Thus the equation is a natural consequence of the stability analysis and the POD of the governing PDE. A mechanical analogue of (\ref{model}) is a damped particle moving on a curved surface forced in the direction $V$ with a velocity $q$ but free in the orthogonal direction $X$. To ensure that the potential characterizes the drop close to its stationary solution the spectrum of our dynamical system (\ref{model}) must approximate that obtained from the linear stability analysis of the stationary pendant drop described in \S3. This leads to the following approximations for the functional form $a,b,\eta$ for $V\le V_{c}$ : 

\begin{equation}
\left\{
\begin{array}{lcl}
a(V) & = &\frac{V^{2} - V_{c}^{2}}{V} \\
b(V)& = & \frac{\alpha}{V^{3}} \\
\eta(V)&=&-2\beta  - \frac{2 \gamma }{\delta + V}
\end{array}
\right.
\label{coef}
\end{equation}

where $\alpha,\beta,\gamma$ and $\delta$ are fitting parameters computed to match the spectrum of the (\ref{disc}). The fitting functional form for (\ref{coef}) of  $a,b,\eta$ are chosen to be simple and to have a minimum of parameters.  However other functional forms do not change the qualitative behaviour of the solution to (\ref{model}). The state variable $X$ eventually diverges when it goes over the unstable branch $U$ or when $V>V_{c}$ which event corresponds to the ejection of a drop. The system is then  reset bringing back $X$ and $V$ close to a stable solution in the same way as the drop goes back to a volume $V<V_c$ and close to a stable drop shape after pinch-off. In Fig. \ref{volvsvel} by solving (\ref{disc}) numerically, we plot the total volume $V$ of the drop at the break-up point as well as the ratio of the ejected volume $V_e$ to the total volume $V$ as functions of the exit velocity $v_{0}$. We find that ejected volume $V_{e}\sim v_{0}V$, so that the volume of the attached drop is a function of the total volume. Therefore, we choose the new value of $X$ for the pendant drop keeping $\partial_{t}X$  the same as at the break-up time; this last reinjection process is arbitrary but does not influence  the results qualitatively.  Numerical simulations of the simple model (\ref{model}-\ref{coef}) plotted in Fig. \ref{simplemodel}b yields a bifurcation diagram for the time between droplets $\tau_{n}$ as a function of $v_0$. Comparing it qualitatively with the same data derived from the hydrodynamic model (\ref{disc}) Fig. \ref{simplemodel}a shows that the transition to chaos occurs in a similar way as the flow rate increases, i.e. $\tau_{n}$ oscillates as a function of $v_{0}$ followed by a simple period-doubling and then by a transition to chaos via a period-doubling bifurcation. When the dripping is chaotic we can also notice several qualitative similarities like a period-3 dripping between two chaotic regime, a reverse period-doubling bifurcation ending in a period-2 dripping regime with a large difference in the interval of time between drops, etc. The surprising similarities between the bifurcation diagrams suggest that simple rationally derived models can capture much of the qualitative dynamics of dripping.   
 
The model (\ref{model},\ref{coef}) is robust with the choice of the functional form $a,b,\eta$ and respect to the reinjection process emphasizing that the qualitative features of the dripping faucet are due to primarily the Saddle-node bifurcation and the damped oscillations leading to it. We note that although our model (\ref{model}-\ref{coef}) may be reminiscent of the widely studied family of mass-spring models ( see  (\cite{Shaw}) as well as  (\cite{kiyono}) and references therein) there is a qualitative difference. Unlike in all previous models, we account for the all important Saddle-Node bifurcation represented by the nonlinear term in the equation (\ref{model})  and thus naturally account for the onset of the necking process.

An even simpler model which ignores the temporal dynamics and focuses exclusively on the interval between drop emissions leads to a return map
\begin{equation}
\left\{
\begin{array}{lcl}
\partial_{t}U&=&(i\omega - \lambda)U, \\
\partial Z & = & \epsilon+Z^{2}. 
\end{array}
\right.
\label{2modemodel}
\end{equation}

\begin{figure}
\begin{center}
\includegraphics[width=12cm]{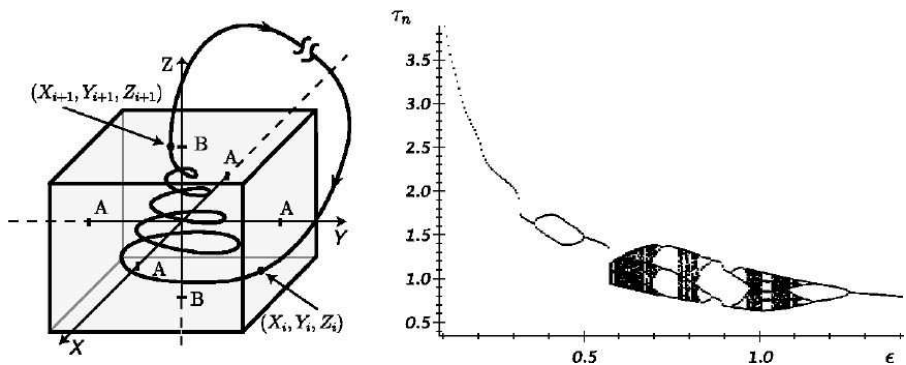}
\caption{Bifurcation diagram for the simple return map (\ref{2modemodel}). $\tau_{n}$ corresponds to the time between two emitted drops and $\epsilon$ to the flow rate (Here $A=1.4, B=1, \lambda=2.6$ and $\omega=20$). Again, we see the similarity to the bifurcation diagram obtained from the complete model (Fig. \ref{simplemodel}a) as well to that obtained from the ODE model (Fig. \ref{simplemodel}b).}
\label{returnmap}
\end{center}
\end{figure}

Here $U=X+iY$ is the amplitude of the oscillatory damped mode with the associated eigenvalue $i\omega-\lambda$ ($\lambda > 0$) and $\epsilon$ characterises the flow rate and is associated with the growing mode $Z$.  As shown in Fig. (\ref{returnmap}) the return map is defined using a parallelopiped of length $(A,A,B)$ in the phase-space $(X,Y,Z)$ centred at the saddle-node point (Fig. (\ref{returnmap}a)). We first contruct the map from the plane $Y=A$, before the Saddle-Node area, to the plane $Z=B$, after the Saddle-Node area. A point $(X_i,A,Z_i)$ is mapped into $(X_{i+1},Y_{i+1},B)$ (see Fig. (\ref{returnmap}a)) via

\begin{equation}
\left\{
\begin{array}{lcl}
X_{i+1}&=&(X_i\cos(\omega\tau_i)-A\sin(\omega\tau_i))e^{-\lambda\tau_i} \\
Y_{i+1}&=&(X_i\cos(\omega\tau_i)+A\sin(\omega\tau_i))e^{-\lambda\tau_i} 
\end{array}
\right.
\label{map}
\end{equation}

where $\tau_i$ is the return time.

\begin{equation}
\tau_i=\frac{\arctan(B/\sqrt{\epsilon})-\arctan(Z_i/\sqrt{\epsilon})}{\sqrt{\epsilon}}
\end{equation}

In order to complete the construction of the dynamical model, we again need a global reinjection process that resets the system and replaces the dynamics of  pinch-off.  The simplest way to model the reinjection flow is via a rigid rotation, as for instance

\begin{equation}
\left\{
\begin{array}{lcl}
X_{i+1}\rightarrow X_{i+1} \\
Y_{i+1}\rightarrow Z_{i+1} 
\end{array}
\right.
\label{reinjection}
\end{equation}

Using (\ref{map}) and (\ref{reinjection}), the Poincar\'e map which models the process is then given by

\begin{equation}
\left\{
\begin{array}{lcl}
\tau_i&=&\displaystyle\frac{\arctan(B/\sqrt{\epsilon})-\arctan(Z_i/\sqrt{\epsilon})}{\sqrt{\epsilon}} \\
X_{i+1}&=&\displaystyle(X_i\cos(\omega\tau_i)-A\sin(\omega\tau_i))e^{-\lambda\tau_i} \\
Z_{i+1}&=&\displaystyle(X_i\cos(\omega\tau_i)+A\sin(\omega\tau_i))e^{-\lambda\tau_i} \label{Pmap}
\end{array}
\right.
\end{equation}

A numerical simulation of this return map shown in Fig. (\ref{returnmap}b) leads to a bifurcation diagram very similar to that of the hydrodynamical model shown in Fig. (\ref{simplemodel}a) and the ODE modes (\ref{simplemodel}b) .

\section{A mechanical analogue of the dripping faucet}
 
\begin{figure}
\begin{center}
\includegraphics[width=13cm]{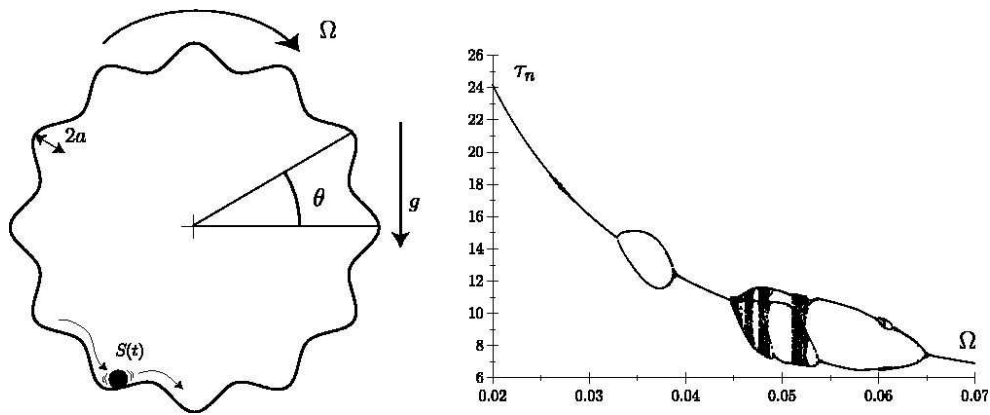}
\caption{(a) A mechanical model for the dripping faucet consists of a damped particle moving inside a cogwheel rotating with an angular velocity $\Omega$. The teeth, i.e. the local minima correspond to the local minima associated with static drops, and the angular velocity of rotation is similar to the flow rate.
(b) The bifurcation diagram for the particle in a cogwheel showing $\tau_n$ the time for the particle to move from one tooth to another as function of $\Omega$ obtained by solving the equation (\ref{mech},\ref{geareq}) with $k=13,\nu=0.25,a=0.01$. Again, note the similarity of this bifurcation diagram to that shown in Fig. \ref{simplemodel},\ref{returnmap}.   }
\label{wheel}
\end{center}
\end{figure}

A natural mechanical analogue of the dripping faucet arises from the analysis presented in the previous sections and is shown in Fig. \ref{wheel}a, where a frictional particle sits inside a corrugated cogwheel rotating about an axis perpendicular to gravity.  The  particle may start at the bottom of one of the local minima, but as the cogwheel rotates, this stable static solution eventually disappears via a  saddle-node bifurcation. The particle then escapes and performs damped oscillations about a new minimum similar to the previous  one, and the scenario is repeated. We see immediately that the rotation rate plays the role of the flow rate in the dripping faucet, while the periodic potential which is locally cubic characterizes the capillary-gravity potential for the drop. The dripping then corresponds to the large excursion of the particle from one minimum to another, accompanied by inertial oscillations and damping.  As the rotation rate becomes similar to the frequency of the particle about its local minima, we can expect complex dynamics here just as in the dripping faucet.

To quantify our mechanical analogue, we use an arc-length coordinate $S(t)$ for the position of the particle. Then its equation of motion, in dimensionless form, is

\begin{equation}
\partial_{tt} S + \nu \partial_t S = -\partial_s y \label{mech}
\end{equation}
Here the vertical position of the particle $y(S(t),t)$ is determined by the instantaneous position of the cogwheel which is given as 
\begin{equation}
\left\{
\begin{array}{lcl}
x(s,t)&=&(1+a\cos k(\theta(s)+\Omega t) )\cos\theta(s) \\
y(s,t)&=&(1+a\cos k(\theta(s)+\Omega t) )\sin\theta(s) 
\label{geareq}
\end{array}
\right.
\end{equation}
As usual, here the relation between the cartesian coordinates and arc-length coordinates is given by $ds=(dx^2+dy^2)^{1/2}$ which leads to
\begin{equation}
ds=\sqrt{1 + 2a\cos{k(\theta(s)+\Omega t)} + a^{2}\cos^{2}k(\theta(s)+\Omega t) + a^{2}k^{2}\sin^{2}k(\theta(s)+\Omega t)}d\theta
\end{equation}
In (\ref{geareq}), $\Omega$ is the angular velocity of the cogwheel, $k$ the wave number of the modulation of the cogwheel $a$ the amplitude of the cogs, and $\nu$ is the dimensionless friction parameter.  When $\Omega=0$ the particle sits in a stable static position in between two teeth.  An unstable solution close to the tip of  the neighboring tooth separates the particle from another static stable solution, and a large enough perturbation will enable the particle to go to  this solution just as a static drop with a volume less than $V_c$ hanging from a faucet with the flow rate $q=0$ will drip if perturbed strongly.  When $\Omega \ne 0$ but is still small, the particle moves from one local minimum to the next, accompanied by damped oscillations at a typical frequency $(\partial_s y)^{1/2}$, which varies with the location of the particle as for the drop. This periodic transition between teeth eventually becomes chaotic when $\Omega$ becomes large enough via a sequence of bifurcations  as shown in 
Fig. \ref{wheel}b. We see that this bifurcation diagram is qualitatively similar to those seen in the dripping faucet shown in Fig. \ref{fullsimu}, and the hierarchy of models considered earlier, i.e. the ODE model bifurcation diagram in Fig. \ref{simplemodel}, and the return map bifurcation diagram shown in Fig. \ref{returnmap}.

\section{Discussion}

Based on the study of the stability of a pendant drop and numerical simulations of a lubrication-type model for the hydrodynamics of a dripping faucet, we have constructed a hierarchy of simple models  that qualitatively account for the various experimentally observed behaviors of a dripping faucet. Following a numerical simulation of the lubrication equations that is used to obtain a bifurcation diagram for the dynamics of dripping, we first show that a simple ODE model based on the dynamics of just two modes is sufficient to capture the qualitative features of the bifurcation diagram. This led us to even simpler model of a discrete map which also shows the same qualitative behaviour. Finally, we propose a new mechanical analogue for the dripping faucet, one whose behaviour is similar to that of the other models. 

All our models differ qualitatively from the many previous models proposed for dripping over the last two decades, by focusing on the  two key elements that govern the dynamics of drop formation and ejection : a) the stability analysis of the static drop close to its critical volume, b) the reinjection of the system back to the neighborhood of a stable static solution. These models are corroborated by a POD of the numerical solution determined by solving (\ref{disc}) and the spectral analysis of the static solution close to the instability threshold. From the perspective of dynamical systems, our models incorporate a coupled system that invokes two classical normal forms associated with the Saddle-Node bifurcation and the Shilnikov mechanism. As we have discussed, the former corresponds to a loss of a stable "Node" solution via the "collision" with an unstable "Saddle" in the context of the linear stability of a drop with $V \leq V_c$. The Shilnikov mechanism for chaos is associated with the presence of damped oscillations and a global reinjection (\cite{holmes2}, \cite{arneodo}) which are physically manifest in the damped capillary oscillations and the pinch-off process which effectively resets the system.

 \begin{figure}
\begin{center}
\includegraphics[width=12cm]{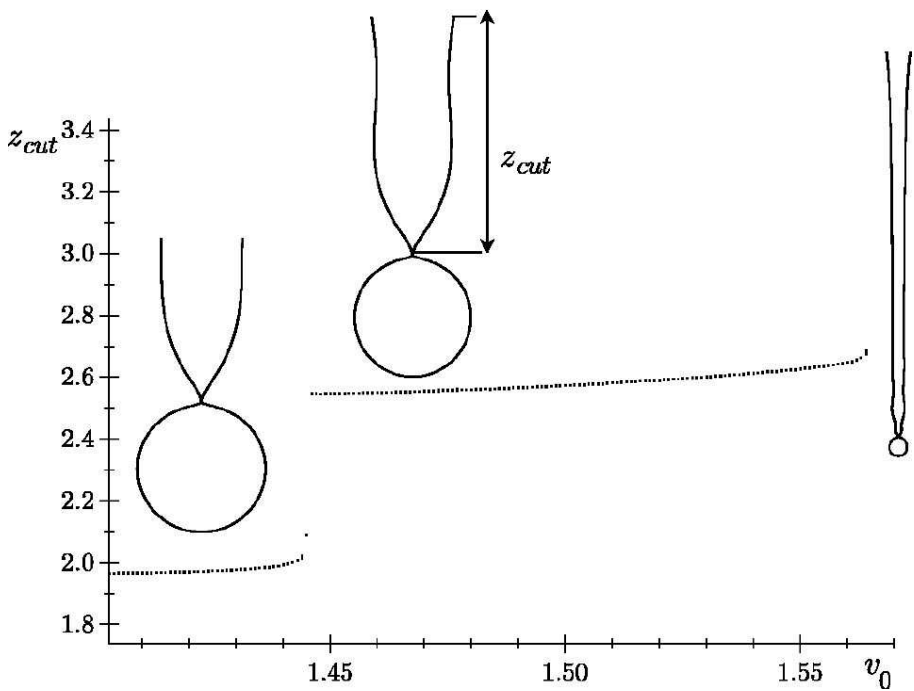}
\caption{Location of the point of drop detachment $z_{cut}$ as a function of the exit velocity $v_0$. We see that as $v_0$ increases, the shape of remaining drop which is originally close to that of the stable solution with no waist, becomes closer to that of the first unstable stationary solution (the drop with one waist, also shown in Fig. 1). Our simple models are clearly not valid in this regime, but by including this mode can be made better. As $v_)$ increases even further, the remaining drop becomes long and slender, thus marking the transition from the dripping to the jetting regime as seen above.}
\label{2drops}
\end{center}
\end{figure}

We conclude by pointing out the limitations and extensions of our simple hydro-dynamical models for dripping. They are valid for small and intermediate flow rates  when the dynamics of dripping is strongly influenced by the static solution, i..e the remainder of the drop after ejection is still similar to the static solution. This notion allows us to characterize and explain the transition from dripping to jetting, observed by \cite{clanet}, \cite{Ambravaneswaran} and others, in a rational way. 
Our simulations of the lubrication equations in \S3 showed that  the necking time $\tau_{n}$ is independent of the exit velocity. During this time, the point where the drop eventually detaches  travels  a distance $l_{d}\sim v_{0}\tau_{n}$ (see \cite{clanet}). As the flow rate increases the length of the remaining drop after ejection increases and so the shape is no longer close to one of the stable stationary solutions shown in Fig.  \ref{volvspb}.  At this stage the dynamics of the dripping is not influenced by the stable stationary solution, an assumption which is the main premise of all our models. This defines the transition from dripping to jetting; in particular,  the influence of the faucet on the dynamics is negligible in this regime as the stable static solution ceases to influence  the dynamical behavior of the system.  The long slender fluid filament in the jetting regime also eventually breaks into drops through the Savart-Plateau-Rayleigh instability, a regime that is qualitatively different from the one we have treated here.

More generally, our approach should be applicable to a variety of systems where we see two relatively common ingredients: the loss of a static stable solution via a saddle-node bifurcation and the dynamical return to a nearby state via a damped oscillatory mode. Examples that come to mind readily include the chaotic nucleation of plumes in convection, frictional oscillations at a rough interface etc.

\end{document}